\documentclass[twocolumn,epjc3]{svjour3}
\smartqed  
\RequirePackage{graphicx}
\journalname{Eur. Phys. J. C}
\begin{document}

\title{Strange stars in Krori-Barua space-time}


\author{{\large Farook Rahaman$^{\ast}$ \thanks{rahaman@iucaa.ernet.in}, Ranjan Sharma$^{\dag}$\thanks{rsharma@iucaa.ernet.in},  Saibal Ray$^{\ddag}$
\thanks{saibal@iucaa.ernet.in}, Raju Maulick$^{\S}$\thanks{rajuspinor@gmail.com} and  Indrani Karar$^{\clubsuit}$\thanks{indrani.karar08@gmail.com}}
\and $^{\ast}${\small Department of Mathematics, Jadavpur
University, Kolkata - 700032, India } \and $^{\dag}$ {\small
Department of Physics, P. D. Women's College, Jalpaiguri 735101,
India} \and $^{\ddag}${\small Department of Physics, Government
College of Engineering \& Ceramic Technology, Kolkata 700 010,
West Bengal, India } \and $^{\S}${\small Department of
Mathematics, Jadavpur University, Kolkata 700 032, West Bengal,
India} \and $^{\clubsuit}${\small Department of Mathematics, Saroj
Mohan Institute of Technology, Guptipara, West Bengal, India}}
\authorrunning{Rahaman {\em et al}}

\date{Received: date / Accepted: date}

\maketitle

\begin{abstract}
The singularity space-time metric obtained by Krori and Barua\cite{Krori1975} satisfies the physical requirements of a realistic star. Consequently, we explore the possibility of applying the Krori and Barua model to describe ultra-compact objects like strange stars. For it to become a viable model for strange stars, bounds on the model parameters have been obtained. Consequences of a mathematical description to model strange stars have been analyzed.
\end{abstract}

\section{Introduction}\label{sec:1}
In relativistic astrophysics, understanding the nature and exact composition of a specific class of compact stars which are more compact than ordinary neutron stars, has become a field of active research in recent years. A neutron star is the final stage of a gravitationally collapsed star which, after exhausting all its thermo-nuclear fuel, gets stabilized by degenerate neutron pressure. Soon after the discovery of the particle `neutron' by Chadwick, the existence of neutron star was predicted. Later on, the concept got observational support with the discovery of pulsars\cite{Hewish}. With the progress in our understanding of particle interactions at high energy, theoretical modelling of neutron stars have improved considerably over the last few decades\cite{Shapiro}. However, the nature of particle interactions beyond nuclear density is still poorly understood. The conjecture that quark matter might be the true ground state of hadrons\cite{Witten,Farhi}, has led to the discussions of an entirely new class of stellar bodies composed of deconfined $u$, $d$ and $s$ quarks, called `strange stars'. It is interesting to note that a strange matter equation of state (EOS) seems to explain the observed compactness of many astrophysical objects like Her X-1, 4U 1820-30, SAX J 1808.4-3658, 4U 1728-34, PSR 0943+10 and RX J185635-3754, whose estimated compactness, otherwise, cannot be explained in terms of a neutron star EOS\cite{Alcock,Haensel,Weber,Garcia,Rodrigues,Bordbar}. Though many more exotic phases may exist at the interior of such class of stars, in this paper, we shall restrict our discussions to the strange matter EOS only. We shall choose a simple EOS for strange quark matter based on the MIT bag model where the quark confinement is assumed to be caused by a universal pressure $B_g$, called the bag constant\cite{Chodos}. The bag model essentially describes the confinement mechanism of quarks inside hadrons. By imposing the condition that the energy per baryon of strange matter be less than that of the nucleon ($939~$MeV), Farhi and Jaffe\cite{Farhi} have shown that for a stable strange quark matter the bag constant should be $B_g \sim 60~$MeV/fm$^{-3}$.

Once the strange matter EOS is known, one can employ numerical techniques to get an estimate of the gross features of a strange star by integrating the Tolman-Oppenheimer-Volkoff(TOV) equations. From the general relativistic view point, question is, what would be the relevant back ground space-time to model such class of compact stars? In a recent work, Avellar and Horvath\cite{Avellar} have considered a wide set of exact and approximate solutions to model strange stars. The objective of the current investigation is to look for a physically viable analytical model which can describe such class of compact stars. To this end, we explore the possibility of applying the Krori and Barua\cite{Krori1975} (henceforth KB) metric to describe the interior space-time of a strange star. The KB model is singularity free and has earlier been shown to be useful to describe realistic stars\cite{Junevicus1976}. Ivanov\cite{Ivanov2002} has shown that the KB model satisfies the necessary energy conditions of a realistic star. Varela {\em et al}\cite{Varela2010}, for a Einstein-Maxwell system, have used the KB model to describe a self-gravitating, charged, anisotropic fluid sphere satisfying a linear and/or non-linear EOS. The key observations made by them are as follows: (i) spheres with vanishing net charge contain fluid elements with unbounded proper charge density located at the fluid-vacuum interface; (ii) inward-directed fluid forces caused by pressure anisotropy may allow equilibrium configurations with larger net charges and electric field intensities than those found in studies of charged isotropic fluids; (iii) possible applications of the model to describe charged strange quark stars, dark matter distributions and massive charged particles. In a separate paper, Farook {\em et al}\cite{Rahaman2010} have used the KB model to analyze an anisotropic, charged, static, spherically symmetric fluid source. It has been shown that the inclusion of a tangential pressure-like variable admits a non-linear, Chaplygin-type EOS. Interestingly, the two approaches coincide for an EOS of the form $p= H\rho$, where $\rho$ is the energy density, $p$ the pressure and $H$ is a model parameter describing the stiffness of the EOS. The results obtained by Varela {\em et al} may be regained by Farook {\em et al}'s\cite{Rahaman2010} approach. They too predicted a possible extrapolation of the investigation to the case of astrophysical bodies, in particular, for a quark star of radius $\sim 8~$ km.

The present investigation is a follow up of the earlier works done by Varela {\em et al}\cite{Varela2010} and Rahaman {\em et al}\cite{Rahaman2010}. In the present work, we shall incorporate the bag model EOS in the KB model and study the subsequent stellar configurations. It is to be noted here that once we specify the EOS, we can integrate the TOV equations to derive the gross features of a stellar configuration. On the other hand, if one of the metric functions is assumed a priori, one can determine the subsequent EOS of the material composition of the star. However, if both the metric functions as well as the EOS are provided, it becomes an over determined system. To overcome the situation, we include two additional input parameters into the system. Note that the KB model was originally developed for a charged neutral, isotropic and spherically symmetric object in static equilibrium. We assume here that the composition of the strange star is anisotropic in nature coupled with high electric field. These assumptions are justified due to the following reasons: Strange stars are extremely dense objects and at very high density, it is expected that the pressure should be anisotropic in nature\cite{Bowers}. The electric field at the surface of a strange star has also been reported to be very high\cite{Usov}. Implications of these additional parameters on the physical behaviour of the strange star will be discussed in the following sections.

The paper has been organized as follows: In Sect.~\ref{sec:2}, we have written the basic field equations. In different sub-sections of Sect.~\ref{sec:3}, we have derived bounds on the model parameters based on various physical requirements. In Sect.~\ref{sec:4}, we have discussed implications of applying the KB model for the description of strange stars. In different sub-sections of Sect.~\ref{sec:5}, we have analyzed various features of the model including stability of the resultant configurations. Finally, some concluding remarks have been made in Sect.~\ref{sec:6}.

\section{Basic Equations}\label{sec:2}
We assume that the interior space-time of a `strange star' is well described by the Krori and Barua\cite{Krori1975} metric given by
\begin{equation}
ds^2 = -e^{\nu(r)}dt^2 + e^{\lambda(r)}dr^2 +r^2(d\theta^2 +sin^2\theta d\phi^2),\label{eq1}
\end{equation}
where, $\lambda(r)=Ar^2$ and $\nu(r) = Br^2 + C$. In Eq.~(\ref{eq1}), $A$, $B$ and $C$ are arbitrary constants which will be fixed on the ground of various physical requirements.

For a static charged fluid source with density $\rho(r)$, radial pressure $p_r(r)$, tangential pressure $p_t(r)$, proper charge
density $\sigma(r)$ and electric field $E(r)$, the Einstein-Maxwell(EM) equations take the form (we employ the geometrized units $G = c = 1$)
\begin{eqnarray}
8\pi\rho+E^2 &=& e^{-\lambda}\left(\frac{\lambda^\prime}{r}-\frac{1}{r^2}\right)+\frac{1}{r^2},\label{eq2}\\
8\pi p_r-E^2 &=& e^{-\lambda}\left(\frac{\nu^\prime}{r}+\frac{1}{r^2}\right)-\frac{1}{r^2},\label{eq3}\\
8\pi p_t+E^2 &=& \frac{e^{-\lambda}}{2}\left(\nu^{\prime\prime}+\frac{\nu^{\prime 2}}{2}+\frac{\nu^{\prime}
-{\lambda^{\prime}}}{r}-\frac{\nu^{\prime}\lambda^{\prime}}{2}\right), \label{eq4}\\
E(r) &=& \frac{1}{r^2}\int_0^r 4\pi r^2 \sigma
e^{\frac{\lambda}{2}}dr = \frac{q(r)}{r^2}, \label{eq5}
\end{eqnarray}
where $q(r)$ is the total charge within a sphere of radius $r$.

Following the MIT bag model, we take the simple form of the strange matter EOS
\begin{equation}
p_r = \frac{1}{3} (\rho -4B_g), \label{eq6}
\end{equation}
where, $B_g$ is the bag constant. With the choice of the above EOS, we have a system of five independent equations with five unknown parameters namely, $\rho$, $p_r$, $p_t$, $E(r)$ and $\sigma(r)$. Substituting the metric potentials given by $\lambda(r)=A r^2$, $\nu(r) = B r^2 + C$, and their derivatives in Eqs.~(\ref{eq2})-(\ref{eq6}), we obtain
\begin{eqnarray}
\rho &=& \frac{3}{ 16 \pi}(A+B)e^{-Ar^2} + B_g,\label{eq7}\\
p_r &=& \frac{ 1}{ 16 \pi}(A+B)e^{-Ar^2} -  B_g,\label{eq8}\\
p_t  &=& \frac{ 1}{ 8\pi}\left[ \left(\frac{7}{2}B-
\frac{3}{2}A +B^2r^2-ABr^2+\frac{1}{r^2} \right)e^{-Ar^2}\right.\nonumber \\
& &\left. - \frac{1}{r^2}\right] + B_g,\label{eq9}\\
E^2 &=& \frac{ 1}{2}\left(A-3B-\frac{2}{r^2}\right)e^{-Ar^2} + \frac{1}{r^2}- 8 \pi B_g.\label{eq10}
\end{eqnarray}
The charge density is obtained as
\begin{eqnarray}
\sigma  &=& \frac{ e^{-\frac{Ar^2}{2}}}{2\pi r  }\psi +
\frac{ Ae^{-\frac{3Ar^2}{2}}}{8\pi r\psi}\left[2-(A -3B)r^2
\right] \nonumber \\
&& + \frac{e^{-\frac{Ar^2}{2}}}{4\pi r^3  \psi}[
e^{-Ar^2}-1],\label{eq11}
\end{eqnarray}
where
$$ \psi =  \sqrt{~\left[ \frac{ 1}{
2}\left(A-3B-\frac{2}{r^2}\right)e^{-Ar^2} + \frac{1}{r^2}- 8 \pi
B_g\right]}. $$
The charge within a sphere of radius $r$ turns out to be
\begin{equation}
q  = r^2\sqrt{\left[\frac{ 1}{
2}\left(A-3B-\frac{2}{r^2}\right)e^{-Ar^2} + \frac{1}{r^2}- 8 \pi
B_g\right]}.\label{eq12}
\end{equation}
The anisotropic stress is obtained as
\begin{eqnarray}
\Delta &=& p_t-p_r = 2B_g -\frac{1}{8\pi r^2} \nonumber \\
&& +\frac{ 1}{8\pi}\left(3B - 2A + B^2 r^2 -A B r^2 +\frac{1}{r^2}
\right)e^{-Ar^2}.\label{eq13}
\end{eqnarray}

\section{Bounds on the model parameters}\label{sec:3}
One of the advantages of using the KB metric is that there there is no singularity in its metric functions. Proper bounds, however, should be imposed on the conatants appearing in the metric functions so that all the physically significant parameters remain well behaved at all interior points of the star.
\subsection{Regularity at the centre ($r=0$):}\label{sec:3.1}
From Eq.~(\ref{eq7}), we obtain the central density in the form
\begin{equation}
\rho_0 = \rho(r=0) = \frac{3}{ 16 \pi}(A+B) + B_g.\label{eq14}
\end{equation}
For regularity of the electric field, it must vanish at the centre, i.e.,
\begin{equation}
E^2(r=0) =  \frac{3}{ 2}(A-B)- 8\pi B_g = 0,\label{eq15}
\end{equation}
which yields
\begin{equation}
B_g = \frac{3}{16\pi}(A-B).\label{eq16}
\end{equation}
Substituting the value of $B_g$ in Eq.~(\ref{eq14}), we note that the parameter $A$ corresponds to the central density
given by
\begin{equation}
A = \frac{8\pi \rho_0}{3}.\label{eq17}
\end{equation}
Eq.~(\ref{eq17}) implies that $A$ is finite and positive. From Eq.~(\ref{eq16}), it then follows that for a positive value of the Bag constant, we must have $A > B$.

The two pressures and density should be decreasing functions of $r$. In our model, radial variation $p_r$ is obtained as
\begin{equation}
\frac{d p_r}{dr}   = -\frac{ 1}{ 8 \pi}(A+B)rAe^{-Ar^2}.\label{eq18}
\end{equation}
Obviously, at $r=0$, $\frac{d p_r}{dr}$ = 0. Now, $\frac{d^2 p_r}{dr^2} < 0$, if the condition $- A(A+B) < 0$ is satisfied.
Since $A$ is positive and $A > B$, this implies that $B > 0$. In Sect.~\ref{sec:5.2}, we have shown that if the strong energy condition has to be satisfied then $B >0$. Therefore, if we assume that the matter within the sphere satisfies the strong energy condition so that $B>0$, then $p_r$ will decrease radially outward. Similarly, it can be shown that $p_t$ also decreases radially outward. The radial variation of the energy density is obtained as
\begin{equation}
\frac{d \rho}{dr} = -\frac{ 3}{ 8\pi}(A+B)rAe^{-Ar^2},\label{eq19}
\end{equation}
which also shows that at $r=0$, $\frac{d \rho}{dr} $ = 0 and
$\frac{d^2\rho}{dr^2} = -\frac{ 3}{ 8 \pi}(A+B) < 0$.
Thus, in this set up, a necessary and sufficient condition for regular behaviour of the physical parameters will be $A > B > 0$.

\subsection{Regularity at the boundary ($r=R$):}\label{sec:3.2}
The radius $R$ of the star can be obtained by utilizing the condition that the radial pressure should vanish at the surface i.e.,
\begin{equation}
p_r ( r=R) = \frac{1}{ 16 \pi}(A+B)e^{-AR^2}- B_g  = 0.\label{eq20}
\end{equation}
This yields
\begin{equation}
R = \sqrt{\frac{1}{A}ln\left[\frac{A+B}{16\pi B_g}\right]}.\label{eq21}
\end{equation}
Since all the parameters on the right hand side of Eq.~(\ref{eq21}) have positive values as discussed in Sect.~\ref{sec:3.1}, the radius of the star is finite and positive.

The exterior space-time of the star will be described by the Reissner-Nordstr\"om metric\cite{Reissner1916,Nordstrom1918} given by
\begin{eqnarray}
ds^{2} &=& -\left(1 - \frac{2M}{r} + \frac {Q^2}{r^2}\right)dt^2 + \left(1 - \frac{2M}{r} + \frac {Q^2}{r^2}\right)^{-1}dr^2
\nonumber\\
&& + r^2(d\theta^2+\sin^2\theta d\phi^2), \label{eq22}
\end{eqnarray}
where, $Q$ is the total charge enclosed within the boundary $r=R$. Continuity of the metric coefficients $g_{tt}$, $g_{rr}$ and
$\frac{\partial g_{tt}}{\partial r}$ across the boundary surface $r= R$ between the interior and the exterior regions of the star yields
the following results:
\begin{eqnarray}
1 - \frac{2M}{R} + \frac {Q^2}{R^2} &=& e^{BR^2+C},\label{eq23}\\
1 - \frac{2M}{R} + \frac {Q^2}{R^2} &=& e^{-AR^2},\label{eq24}\\
\frac{M}{R^2} - \frac {Q^2}{R^3} &=& B Re^{BR^2+C}.\label{eq25}
\end{eqnarray}
Eqs.~(\ref{eq23}) - (\ref{eq25}) determine the values of the constants $A$, $B$ and $C$ in terms of the total mass $M$, radius
$R$ and charge $Q$. By solving the above set of equations, we get
\begin{eqnarray}
A &=& - \frac{1}{R^2} \ln \left[ 1 - \frac{2M}{R} + \frac {Q^2}{R^2}
\right], \label{eq26}\\
B &=& \frac{1}{R^2} \left[\frac{M}{R} - \frac {Q^2}{R^2}\right] \left[1 - \frac{2M}{R} + \frac {Q^2}{R^2}
\right]^{-1},\label{eq27}\\
C &=&  \ln \left[ 1 - \frac{2M}{R} +
\frac {Q^2}{R^2} \right]- \frac{ \frac{M}{R} - \frac {Q^2}{R^2}}{
\left[ 1 - \frac{2M}{R} + \frac {Q^2}{R^2} \right]}. \label{eq28}
\end{eqnarray}
Note that the values of the parameters $M$, $R$ and $Q$ should be such that the condition $A > B > 0$ is satisfied. Moreover, consistency of the above equations puts the following constraint on the system:
\begin{eqnarray}
\left[\frac{M}{R} - \frac {Q^2}{R^2}\right] \left[1 - \frac{2M}{R}
+  \frac {Q^2}{R^2} \right]^{-1} \nonumber \\
+\left[\frac{2+ \frac{2M}{R} -
\frac {Q^2}{R^2}}{4 - \frac{2M}{R} + \frac {Q^2}{R^2}}\right]
\times \ln \left[1 - \frac{2M}{R} + \frac
{Q^2}{R^2}\right] = 0.\label{eq29}
\end{eqnarray}
Eq.~(\ref{eq29}) is useful to get an estimate of the charge to radius ($Q/R$) ratio for a given compactness ($M/R$) of the star. Based on logarithmic principle, another condition that must be fulfilled is that the total charge $Q^2 < 2R M$. Therefore, physical values of the parameters like the mass, size and charge can not be fixed arbitrarily in this construction. In Sect.~\ref{sec:4}, we have demonstrated that it is possible to choose numerical values of masses and radii, consistent with the above constraints. These physical values indicate that the corresponding configurations are more likely to be strange stars rather than neutron stars.

\section{Estimation of physical values}\label{sec:4}
\begin{figure}[htbp]
\centering
\includegraphics[scale=.3]{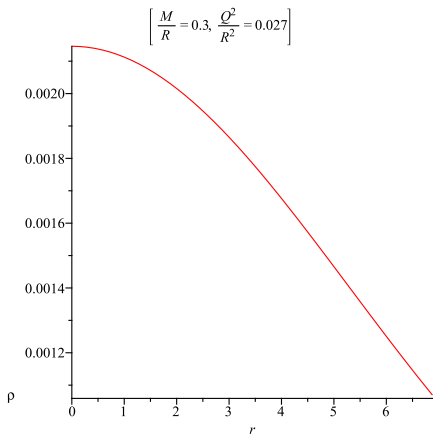}
\caption{The energy density ($\rho$) plotted against $r$. \label{fig1}}
\end{figure}

\begin{figure}[htbp]
\centering
\includegraphics[scale=.3]{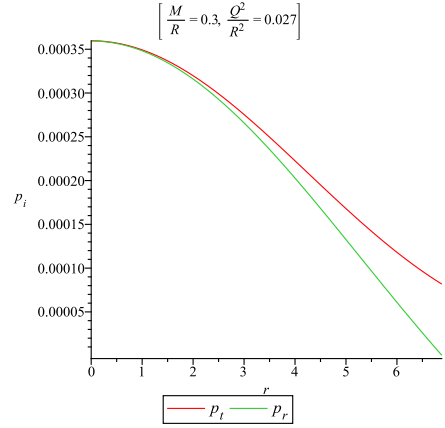} \caption{Radial ($p_r$) and transverse ($p_t$) pressures plotted
against $r$. \label{fig2}}
\end{figure}

\begin{figure}[htbp]
\centering
\includegraphics[scale=.3]{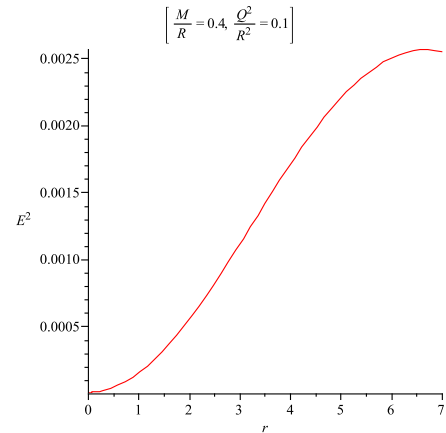} \caption{The electric field ($E^2$) plotted against $r$. \label{fig3}}
\end{figure}

In this section, we assume the mass and radius of a star consistent with the bounds discussed in the previous section so that the compactness the star is greater than that of a neutron star. This will, in turn, help us to get an estimate of the physically relevant parameters like the energy density, pressure and the bag constant. We have considered compact stars of different compactification factors and calculated the corresponding constants. The results have been shown in Table~\ref{tab:1} \& \ref{tab:2}. For example, for star of mass $1.4~M_{\odot}$ and radius $R=6.88~$km, the values of the constants are obtained as $A=0.017977861$,
$B=0.013506968$, $B_g = 0.0002669721$ in units of km$^{-2}$ and $Q^2/R^2 =0.027$ (Case II). Plugging in $G$ and $c$ in the
relevant equations, the values turn out to be $\rho_0 = 2.895\times 10^{15}$ gm~cm$^{-3}$, $\rho_R = 1.443\times 10^{15}$ gm~cm$^{-3}$, $p_r(r=0) = p_t(r=0) = 4.361\times 10^{35}$ dyne~cm$^{-2}$ and the bag constant $B_g = 202.275$ MeV~fm$^{-3}$. Numerical values for other cases have been shown in Table~\ref{tab:3}. Note that each case satisfies the condition $A > B > 0$. Making use of the constraint Eq.~(\ref{eq29}), we also note that for $Q^2/R^2 = 0.004$, the minimum value of $M/R = 0.25$. Therefore, for a star of mass $1.4~M_{\odot}$, the corresponds maximum radius turns out to be $R=8.26~$km. To illustrate the behaviour of physical parameters at the interior of the star, we have considered case II and plotted the variations of the energy density and the two pressures in Fi.~\ref{fig1} - \ref{fig3}, respectively.

The bag constant, in this framework, increases with the increase of the compactification factor, i.e., the bag constant is density dependent. A more dense star requires a greater bag constant. Similar observations may be found in Ref.~\cite{Bordbar2}, where a density dependent bag constant has been employed to model magnetized strange quark stars. The pioneering works of Farhi and Jaffe\cite{Farhi} showed that for a stable strange matter distribution, the bag constant should be around $\sim 60~$Mev/{fm}$^3$. Our results show that with the inclusion of anisotropy and electric field, the bag constant turns out to much more than its representative value. However, we would like to point out here that in Ref.~\cite{Farhi}, the calculation was made for a $\beta$-equilibrium strange quark matter satisfying the baryon number conservation principle where the charge neutrality condition was employed. The window of stability was parametrized by three factors namely, the bag constant, the mass of the quark particles and the QCD coupling constant. What happens when the charge neutrality condition is not imposed is not obvious from the analysis. For a relatively massive strange quark mass there could be an accumulation of net positive charge within the system. Probably, in the presence of charge, to counter the repulsive force generated due to the electric field, the bag pressure increases. The issue, however, is a matter of further investigation. What we have shown here is that if one wishes to use a mathematically consistent and physically reasonable analytic solution to model strange stars, the bag constant does not remain constant. Rather it becomes a free paramter which depends on the compactness of the star.

\begin{table}
\caption{Values of $Q^2/R^2$ for different choices of the compactification factor $M/R$.} \label{tab:1}
\begin{tabular}{@{}lllll@{}}
\hline
Case & $M$ ($M_{\odot}$) & $R$ (km) & $\frac{M}{R}$ & $\frac{Q^2}{R^2}$\\
\hline
I & 1.4 & 8.26 & 0.25 & 0.004 \\
II & 1.4 & 6.88 & 0.30 & 0.027  \\
III & 1.4 & 5.90 & 0.35 & 0.061  \\
IV & 1.4 & 5.16 & 0.40 & 0.105 \\
\hline
\end{tabular}
\end{table}

\begin{table}
\caption{Values of the model parameters $A$ and $B$ as well as the bag constant $B_g$ for different choices of
compactification factor $M/R$. Data obtained in Case II have been utilized to plot the figures.}
\label{tab:2}
\begin{tabular}{@{}llll@{}}
\hline
Case & $A$ (km$^{-2}$) & $B$ (km$^{-2}$) & $B_g$ (km$^{-2}$) \\
\hline
I & 0.0102 & 0.0073 & 0.0001732 \\
II & 0.01798 & 0.01351 & 0.000267 \\
III &  0.0292 & 0.0231 & 0.0003643 \\
IV & 0.044 & 0.037 & 0.000418\\ \hline
\end{tabular}
\end{table}

\begin{table}
\caption{Energy density, pressure and bag
constant for different cases shown in Table~\ref{tab:1} \& \ref{tab:2}.}
\label{tab:3}
\begin{tabular}{@{}llllll@{}}
\hline
Case & $\rho (r=0)$ & $\rho (r=R)$ & $p_r(r=0)$ & $B_g$\\
 & (gm~cm$^{-3}$) & (gm~cm$^{-3}$) & (dyne~cm$^{-2}$) & (MeV~fm$^{-3}$)\\ \hline
 I & $1.643\times 10^{15}$ & $9.362\times 10^{14}$ & $2.123\times 10^{35}$ & 131.204 \\
 II & $2.895\times 10^{15}$ & $1.443\times 10^{15}$ & $4.361\times 10^{35}$ & 202.275 \\
 III & $4.703\times 10^{15}$ & $2.015\times 10^{15}$ & $8.204\times 10^{35}$ & 275.98 \\
 IV & $7.087\times 10^{15}$ & $2.58\times 10^{15}$ & $1.448\times 10^{35}$ & 316.699 \\
\hline
\end{tabular}
\end{table}

\section{Some salient features of the model}\label{sec:5}
\subsection{Generalized TOV equations}\label{sec:5.1}
\begin{figure}[htbp]
\centering
\includegraphics[scale=.3]{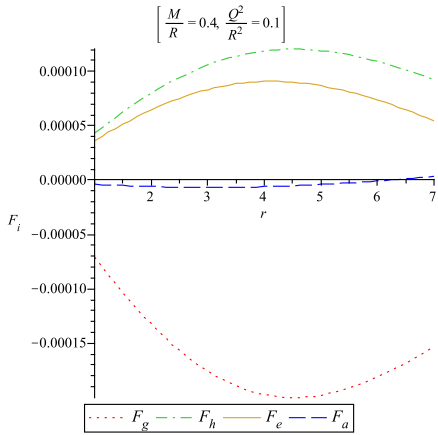} \caption{Contributions of different force terms acting on fluid
elements in static equilibrium. \label{fig4}}
\end{figure}

The  generalized Tolman-Oppenheimer-Volkoff(TOV) equations, in this set up, gets the form
\begin{equation}
-\frac{M_G\left(\rho+p_r\right)}{r^2}e^{\frac{\lambda-\nu}{2}}-\frac{dp_r}{dr}+\sigma
\frac{q
}{r^2}e^{\frac{\lambda}{2}}+\frac{2}{r}\left(p_t-p_r\right) = 0,\label{eq30}
\end{equation}
where $M_G$ is the effective gravitational mass given by
\begin{equation}
M_G(r)=\frac{1}{2}r^2e^{\frac{\nu-\lambda}{2}}\nu^{\prime} = B r^3 e^{\frac{1}{2}[(B-A)r^2 -C]}.\label{eq31}
\end{equation}
Eq.~(\ref{eq30}) describes the equilibrium condition for a charged anisotropic fluid subject to gravitational ($F_g$), hydrostatic
($F_h$), electric ($F_e$) and anisotropic stress ($F_a$) so that
\begin{equation}
F_g+ F_h+ F_e+ F_a=0, \label{eq32}
\end{equation}
where,
\begin{eqnarray}
F_g &=& -B r\left(\rho+p_r\right) = - \frac{Br}{4\pi} (A+B)e^{-Ar^2}, \label{eq33}\\
F_h &=&  -\frac{dp_r}{dr} = \frac{ 1}{ 8 \pi}(A+B)rAe^{-Ar^2},\label{eq34}\\
F_e &=& \sigma E e^{\frac{A r^2}{2}} = \frac{1}{2\pi r}\left[\frac{e^{-A r^2/2}}{2} \left(A-3B-\frac{2}{r^2}\right)\right.\nonumber\\
&&\left. +\frac{1}{r^2} -8\pi B_g\right] + \frac{A e^{-A r^2}}{8\pi r}\left(2-(A-3B)r^2\right) \nonumber\\
&&+\frac{1}{4\pi r^2}\left(e^{-A r^2}-1\right),\label{eq35}\\
F_a &=& \frac{2}{r}\left(p_t-p_r\right) = \frac{2}{r}\left[\frac{1}{8\pi}\left[\left(3B -2A +B^2
r^2 -AB r^2 \right.\right.\right. \nonumber\\
&& \left.\left.\left. +\frac{1}{r^2} \right)e^{-Ar^2} - \frac{1}{r^2}
\right]+2B_g \right]. \label{eq36}
\end{eqnarray}
In Fig.~\ref{fig4}, the profiles of these force terms at the interior of the star for the case II have been shown. The plots indicate that an equilibrium stage can be achieved under the combined effects of gravitational, electric, hydrostatic and anisotropic stresses.

\subsection{Energy conditions}\label{sec:5.2}
\begin{figure}[htbp]
\centering
\includegraphics[scale=.3]{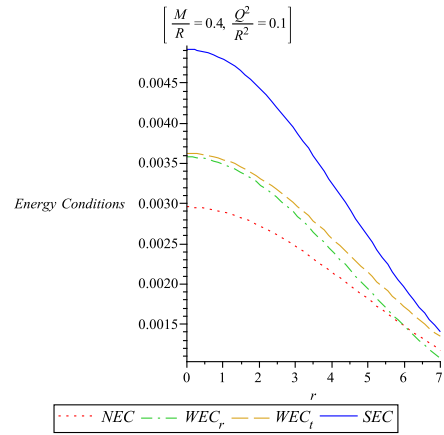} \caption{Different energy conditions plotted against $r$ for case II.} \label{fig5}
\end{figure}

The anisotropic charged fluid sphere composed of strange matter will satisfy the null energy condition (NEC), weak energy condition (WEC) and strong energy condition (SEC) if the following inequalities hold simultaneously at all points within the star:
\begin{eqnarray}
\rho + \frac{ E^2}{8\pi}\geq 0, \label{eq37}\\
\rho + p_r\geq 0, \label{eq38}\\
\rho + p_t + \frac{E^2}{4\pi}\geq 0, \label{eq39}\\
\rho + p_r + 2p_t+\frac{E^2}{4\pi}\geq 0.\label{ec40}
\end{eqnarray}
Employing these energy conditions at the centre ($r=0$), we get the following bounds on the constants $A$ and $B$:\\
(i)  NEC: $\rho + \frac{ E^2}{8\pi}\geq 0 \Rightarrow A \geq 0 $.\\
(ii) WEC: $\rho+p_r\geq 0 \Rightarrow  A +B \geq 0,~~~\rho+p_t+\frac{E^2}{4\pi}\geq 0 \Rightarrow  A +B \geq 0$ \\
(iii) SEC: $ \rho+p_r+ 2p_t+\frac{E^2}{4\pi}\geq 0 \Rightarrow   B \geq 0$.\\
Since the central density is given by  $\rho_0 = \frac{3A}{8\pi}$, we must have $A > 0$, i.e., condition (i) is satisfied.  The weak
and strong energy conditions (ii) and (iii) will then be satisfied if $ B \geq 0$. With the set of values discussed Sect.~\ref{sec:4}, we have shown in Fig.~\ref{fig5}  that the energy conditions are satisfied simultaneously within the sphere.

\subsection{Stability}\label{sec:5.3}
\begin{figure}[htbp]
\centering
\includegraphics[scale=.3]{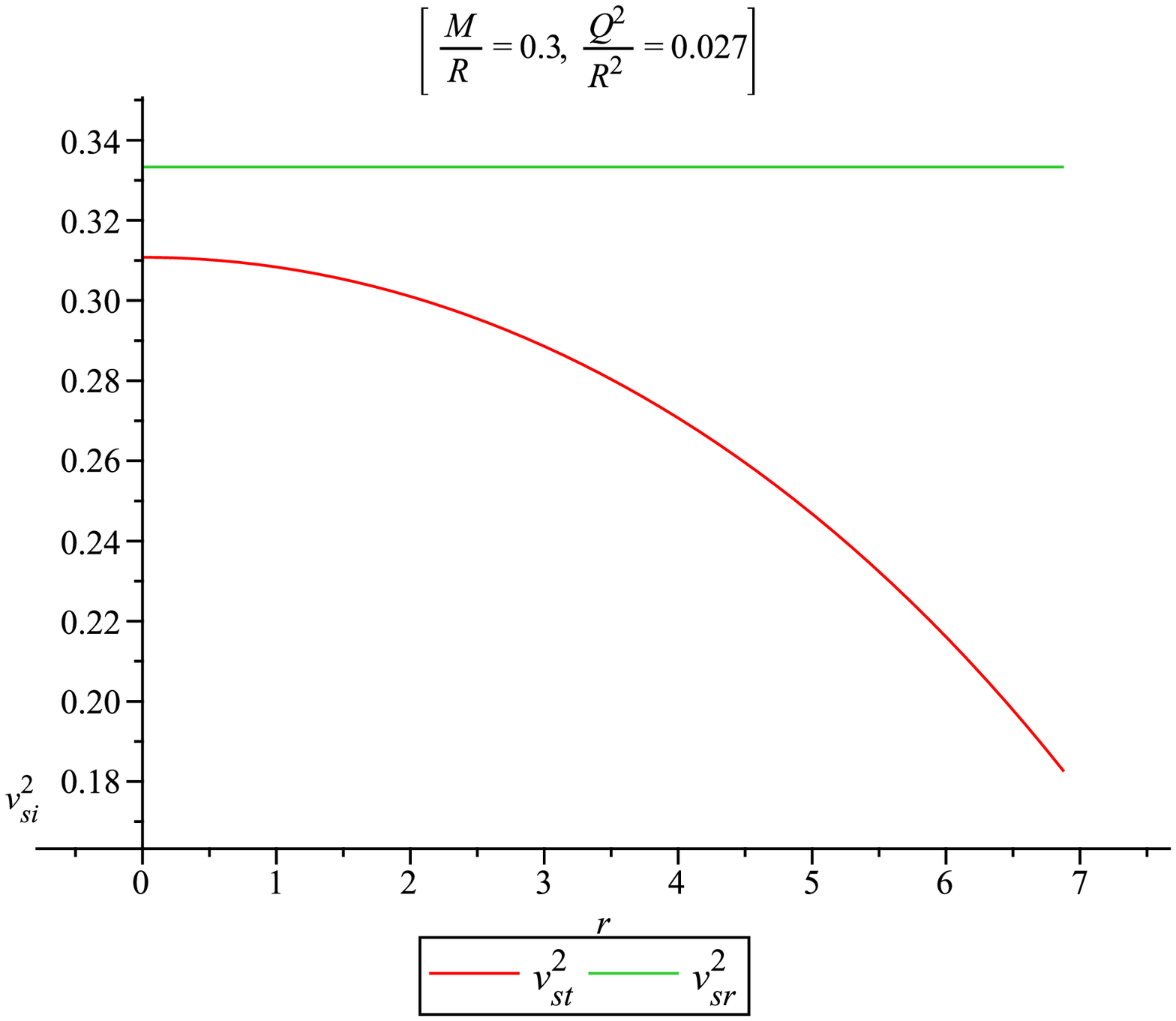} \caption{$\mid v_{st}^2 -
v_{sr}^2 \mid $ plotted against $r$. } \label{fig6}
\end{figure}

\begin{figure}[htbp]
\centering
\includegraphics[scale=.3]{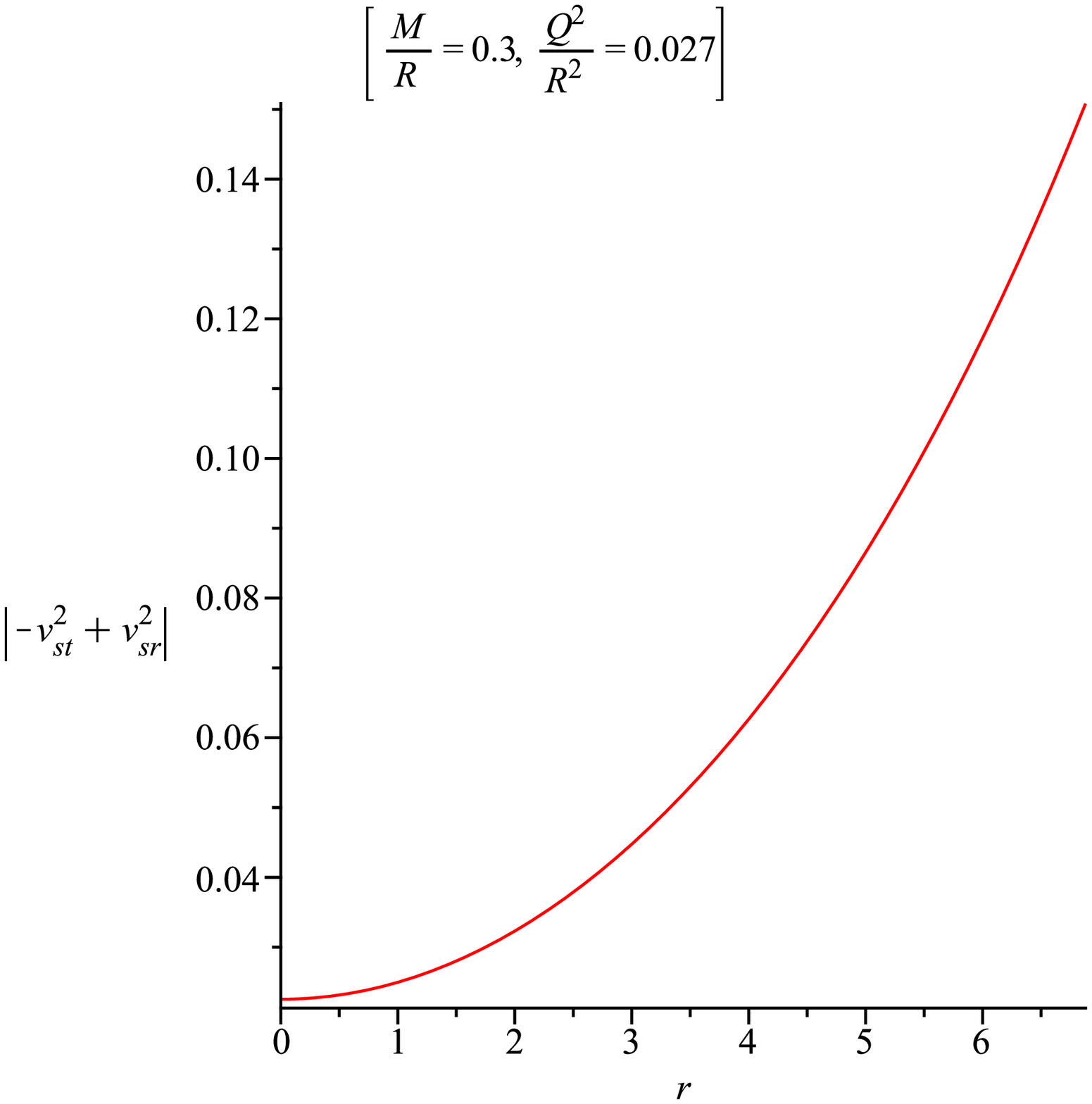} \caption{$\mid v_{st}^2 -
v_{sr}^2 \mid $ plotted against $r$. } \label{fig7}
\end{figure}

\begin{figure}[htbp]
\centering
\includegraphics[scale=.3]{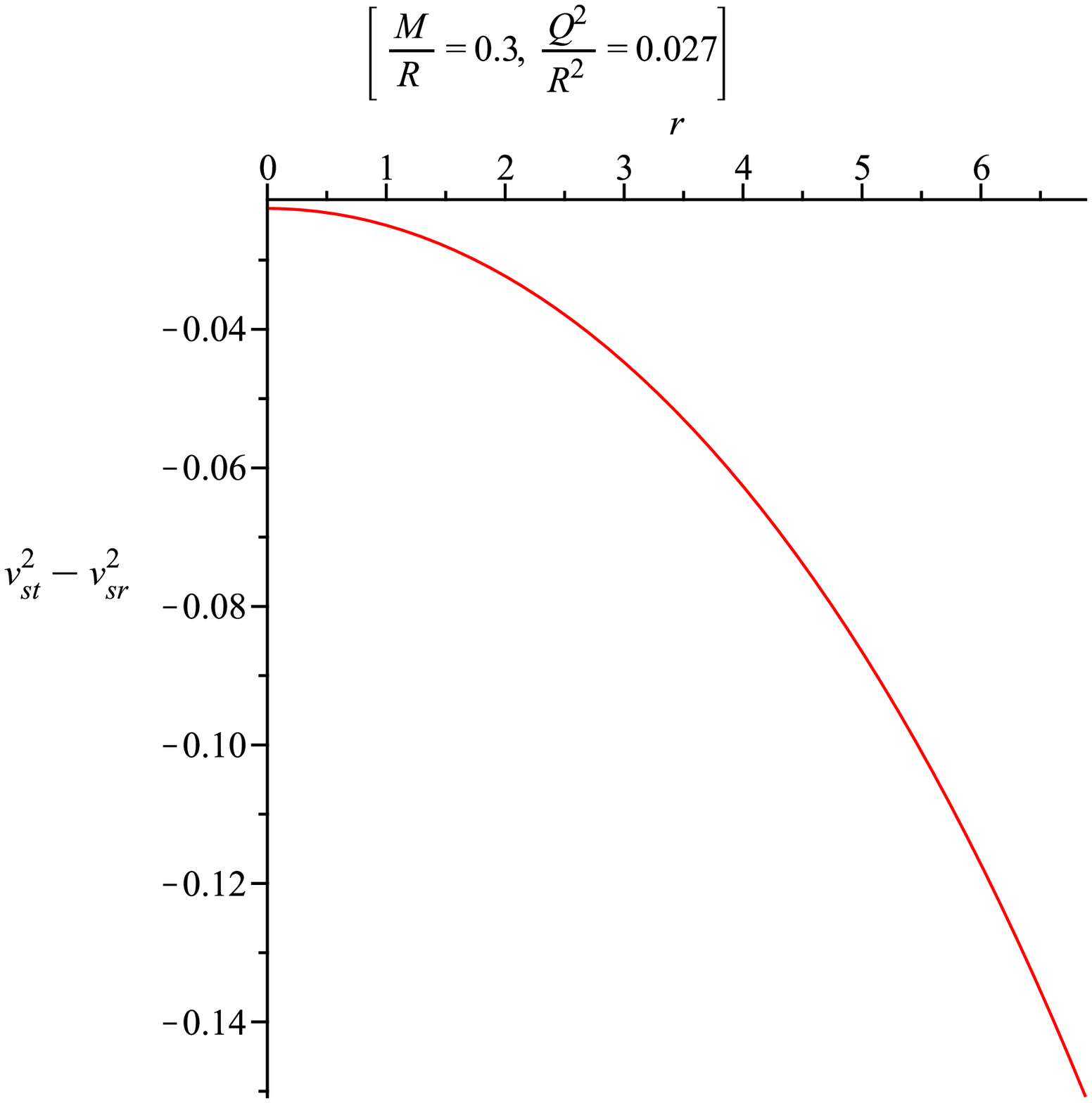} \caption{$ v_{st}^2 - v_{sr}^2$ plotted against $r$. } \label{fig8}
\end{figure}

To examine stability of the resultant configuration, we employ the technique based on Herrera's\cite{Herrera1992} cracking (or overturning) concept. Physical acceptability conditions for a fluid distribution include the condition of causality. It suggests that the squares of the radial and tangential sound speeds should be within the limit $[0,1]$. Herrera's \cite{Herrera1992} cracking (or overturning) concept implies that the region for which the radial speed of sound is greater than that of transverse speed is a
potentially stable region. It also suggests that, for `no cracking' to occur, the difference of the two sound speeds, i.e.,
$v_{st}^2 - v_{sr}^2$ should retain the same sign everywhere within the matter distribution. In our model, we have
\begin{eqnarray}
v_{sr}^2 &=& \frac{dp_r}{d\rho} = \frac{1}{3},\label{eq41}\\
v_{st}^2 &=& \frac{dp_t}{d\rho} = \frac{\alpha +\beta}{-3(A+B)rAe^{-Ar^2}},\label{eq42}
\end{eqnarray}
where,
\begin{eqnarray}
\alpha &=& e^{-Ar^2}\left( 2B^2r -2ABr -  \frac{2}{r^3}\right)+\frac{1}{r^2},\nonumber \\
\beta &=& -2Are^{-Ar^2}\left(\frac{7}{2}B- \frac{3}{2}A +B^2r^2-ABr^2 +\frac{1}{r^2}\right).\nonumber
\end{eqnarray}
For causality condition to be satisfied we must have
\begin{equation}
0 < \frac{\alpha+\beta}{-3(A+B)rAe^{-Ar^2}} < 1. \label{eq43}
\end{equation}
For the assumed set of values, we note that $0\leq v_{sr}^2 \leq 1$ and $0\leq v_{st}^2 \leq 1$ as shown in Fig.~\ref{fig6}.
Following the works of Herrera\cite{Herrera1992} and Andr\'{e}asson\cite{Andreasson1992}, we note that the configuration will be
stable provided $\mid v_{st}^2 - v_{sr}^2 \mid \leq 1 $. Fig.~\ref{fig7}-\ref{fig8} show that the above condition is
satisfied for the assumed configuration implying stability of the configuration.

\subsection{Effective mass-radius relation}\label{sec:5.4}
\begin{figure}[htbp]
\centering
\includegraphics[scale=.3]{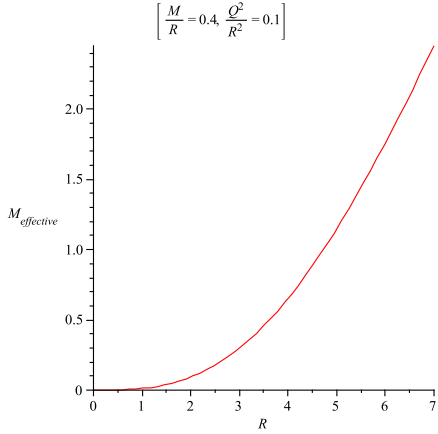} \caption{$M_{eff}$ plotted against
$R$. \label{fig9}}
\end{figure}

\begin{figure}[htbp]
\centering
\includegraphics[scale=.3]{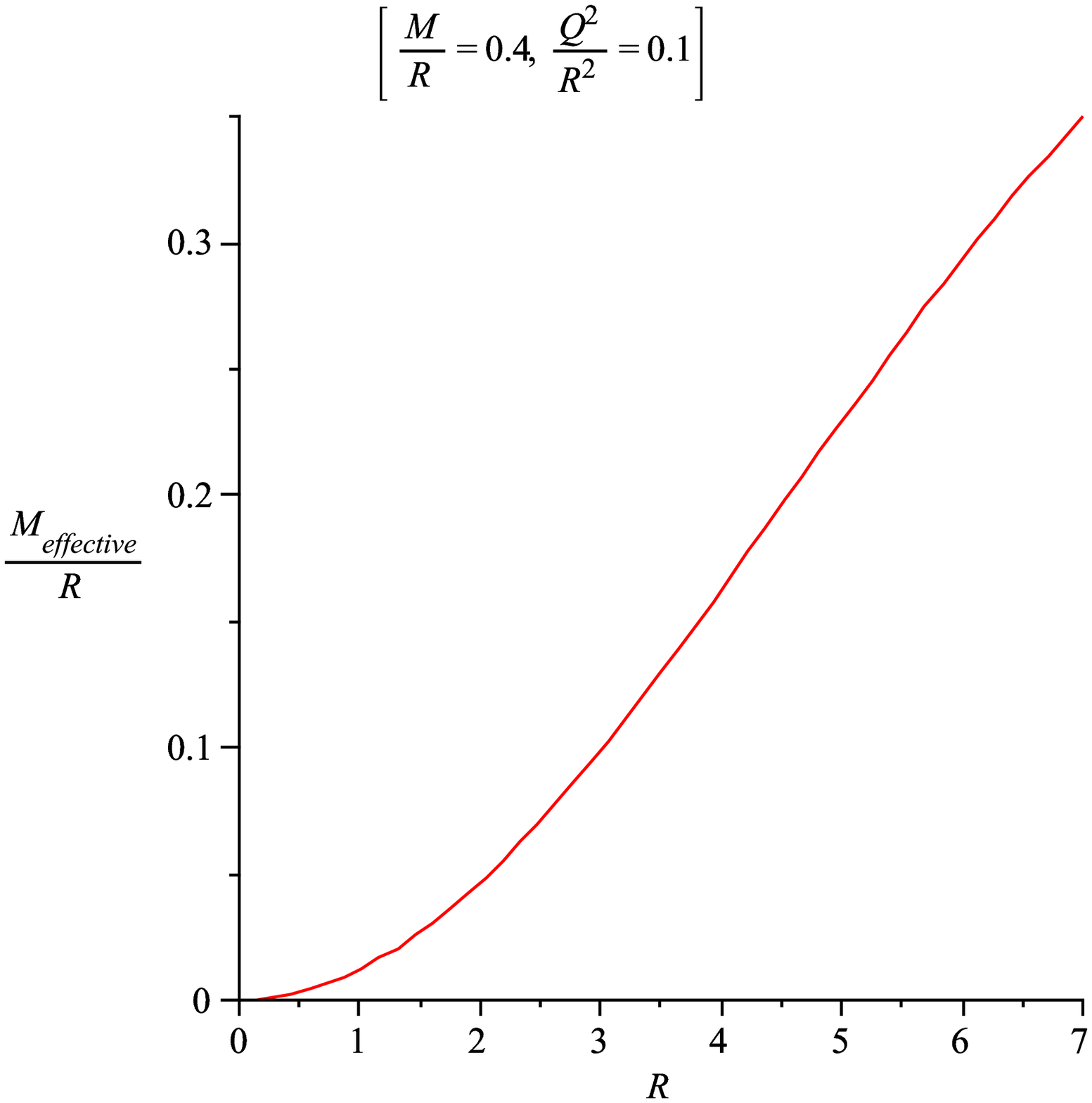} \caption{$\frac{M _{eff}}{R}$ plotted against $R$. \label{fig10}}
\end{figure}

For a static spherically symmetric perfect fluid star, Buchdahl\cite{Buchdahl1959} derived an upper limit for maximum allowed
mass to radius ratio as $\frac{2M}{R} < \frac{8}{9}$ which was later generalized by Mak {\em et al}\cite{Mak2001} for a charged
sphere. In this model, the effective gravitational mass has the form
\begin{equation}
M_{eff} = 4\pi\int^{R}_{0}\left(\rho+\frac{E^2}{8\pi}\right)r^2 dr =
\frac{1}{2}R\left( 1-e^{-AR^2}\right).\label{eq44}
\end{equation}
In Fig.~\ref{fig9}, we have shown the variation of mass against radius. We have also plotted $\frac{M _{eff}}{R}$ against $R$ in Fig.~\ref{fig10} which shows that the ratio $\frac{M _{eff}}{R}$ is an increasing function of the radial parameter. We note that the constraint on the maximum allowed mass-radius ratio in this case is similar to that of an isotropic fluid sphere, i.e., $\frac{M}{R} < \frac{4}{9}$, as obtained by Buchdahl\cite{Buchdahl1959}. Defining the compactification factor as
\begin{equation}
u = \frac{ M_{eff}(R)} {R}=  \frac{1}{2}\left(1-e^{-AR^2} \right),\label{eq45}
\end{equation}
the surface red-shift ($Z_s$) corresponding to the above compactness ($u$) is obtained as
\begin{equation}
Z_s= ( 1-2 u)^{-\frac{1}{2}} - 1=
e^{\frac{AR^2}{2}}-1.\label{eq46}
\end{equation}
The maximum surface redshift, in this set up, for a strange star of mass $1.4~M_{\odot}$ and radius $6.88~$km turns out to be $Z_s = .5303334$.

A lower bound on the mass to radius ratio for a charged sphere has been reported by Andr\'{e}asson\cite{Andreasson1992} which has the form
\begin{equation}
\sqrt{M} < \frac{\sqrt{R}}{3} + \sqrt{\frac{ R }{9} + \frac{Q^2}{3R}}.\label{eq47}
\end{equation}
This inequality is applicable to stellar  objects satisfying the inequality $p_r+ 2p_t  \leq \rho$. In Fig.~\ref{fig12}, we have
plotted $p_r+ 2p_t - \rho$ against $r$ which shows that at all interior points, the above condition is satisfied, i.e.,
Andr\'{e}asson's inequality holds in our model.

\begin{figure}[htbp]
\centering
\includegraphics[scale=.3]{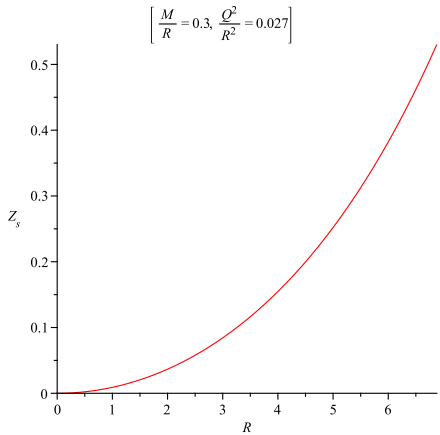} \caption{The redshift function $Z_s$
plotted against $R$. \label{fig11}}
\end{figure}

\begin{figure}[htbp]
\centering
\includegraphics[scale=.3]{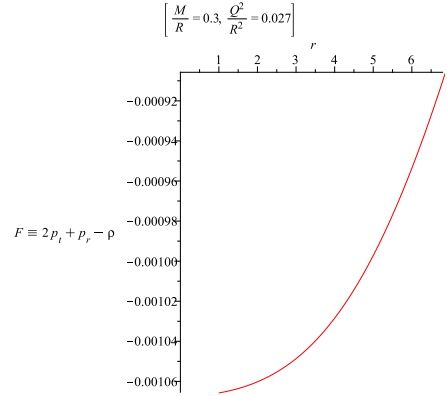} \caption{$2p_t+p_r - \rho$ plotted against $r$. \label{fig12}}
\end{figure}

\section{Discussions}\label{sec:6}
We have explored the relevance of KB model\cite{Krori1975} in the modelling of strange quark stars where the strange matter EOS based on the MIT bag model has been assumed. The inclusion of the EOS does not make the system over determined since the matter distribution in the set up has been assumed to anisotropic in nature together with high electric field. We have shown that a self-consistent mathematical model can generate physical values which are consistent with a strange star. The bag constant, however, in this framework turns out to be above the range specified for a stable strange quark matter, i.e.,  ($60-80~$Mev/fm$^3$)\cite{Farhi,Weber,Alcock}. However, experimental results from CERN-SPS and RHIC show that a wide range of values of the bag constant is possible for a density-dependent bag model\cite{Burgio}. Perhaps, to compensate the extra repulsion in the presence of electric field, the bag value increases in our set up. The impacts of anisotropy and high electric field on the bag model is a matter of further investigation and will be taken up elsewhere.

\begin{acknowledgements}
FR, RS and SR gratefully acknowledge support from the
Inter-University Centre for Astronomy and Astrophysics (IUCAA),
Pune, India, under the  visiting research associateship programme,
under which a part of this work was carried out. FR is also
thankful to PURSE, DST and UGC, Govt. of India,  under Research
Award, for providing financial support.
\end{acknowledgements}



\end{document}